\documentclass[sigconf]{acmart}

\AtBeginDocument{%
  }

\copyrightyear{2026}
\acmYear{2026}
\setcopyright{rightsretained}
\acmConference[CHI EA '26]{Extended Abstracts of the 2026 CHI Conference on Human Factors in Computing Systems}{April 13--17, 2026}{Barcelona, Spain}
\acmBooktitle{Extended Abstracts of the 2026 CHI Conference on Human Factors in Computing Systems (CHI EA '26), April 13--17, 2026, Barcelona, Spain}\acmDOI{10.1145/3772363.3778687}
\acmISBN{979-8-4007-2281-3/2026/04}

\begin{document}

\title[The Future of Cognitive Personal Informatics]{The CHI26 Workshop on the Future of Cognitive Personal Informatics}

\author{Christina Schneegass}
\email{c.schneegass@tudelft.nl}
\orcid{0000-0003-3768-5894}
\affiliation{%
  \institution{Delft University of Technology}
  \city{Delft}
  \country{Netherlands}
}

\author{Francesco Chiossi}
\email{francesco.chiossi@um.ifi.lmu.de}
\orcid{0000-0003-2987-7634}
\affiliation{%
  \institution{LMU Munich}
  \city{Munich}
  \country{Germany}}

\author{Anna L. Cox}
\email{anna.cox@ucl.ac.uk}
\orcid{0000-0003-2231-2964}
\affiliation{%
  \institution{University College London}
  \city{London}
  \country{UK}
}

\author{Dimitra Dritsa}
\email{d.dritsa@tudelft.nl}
\orcid{0000-0002-7615-8520}
\affiliation{%
  \institution{Delft University of Technology}
  \city{Delft}
  \country{Netherlands}
}

\author{Teodora Mitrevska}
\email{teodora.mitrevska@ifi.lmu.de}
\orcid{0009-0005-4263-2510}
\affiliation{%
  \institution{LMU Munich}
  \city{Munich}
  \country{Germany}
  }

\author{Stephen Rainey}
\email{S.Rainey@tudelft.nl}
\orcid{0000-0002-5540-6046}
\affiliation{%
  \institution{Delft University of Technology}
  \city{Delft}
  \country{Netherlands}
}

\author{Max L. Wilson}
\email{max.wilson@nottingham.ac.uk}
\orcid{0000-0002-3515-6633}
\affiliation{%
  \institution{University of Nottingham}
  \city{Nottingham}
  \country{UK}
}

\renewcommand{\shortauthors}{Schneegass et al.}

\begin{abstract}
Research on Cognitive Personal Informatics (CPI) is steadily growing as new wearable cognitive tracking technologies emerge on the consumer market, claiming to measure stress, focus, and other cognitive factors. At the same time, with generative AI offering new ways to analyse, visualize, and interpret cognitive data, we hypothesize that cognitive tracking will soon become as simple as measuring your heart rate during a run. Yet, cognitive data remains inherently more complex, context-dependent, and less well understood than physical activity data. This workshop brings together HCI experts to discuss critical questions, including: How can complex cognitive data be translated into meaningful metrics? How can AI support users' data sensemaking without over-simplifying cognitive insights? How can we design inclusive CPI technologies that consider inter-personal variance and neurodiversity? We will map challenges and opportunities for CPI, considering recent AI advancements, and outline a research road map for the foreseeable future.
\end{abstract}



\maketitle

\section{Introduction \& Motivation}

Quantifying one's life to gain a deeper understanding of one's own body and mind has become a daily routine for many people. ``Personal Informatics'' (PI) \cite{epstein2015lived} systems help visualise data on people's physical activity, sleep, eating, or period, with the goal to support self-monitoring and reflection. In recent years, PI has extended into cognitive data, too, being referred to using the term  ``\textit{Cognitive} Personal Informatics'' (CPI), respectively \cite{schneegass2023future, wilson22sig, wilson2018mental}. Similar to fitness trackers showing users their physical activity, CPI systems support people in reflecting on their cognitive processes (e.g., focus, workload, fatigue, or stress), identifying patterns, improving their well-being, or achieving a personal goal. 

With every year, new consumer neurotechnologies and cognition-aware wearables enter the consumer market, i.e., technologies that non-invasively measure, analyse, and interpret people's cognitive activities. These new devices are smaller, cheaper, more comfortable to wear, and more powerful, and they collect signals that either directly or indirectly provide insights into cognition. Direct measurements, such as Electroencephalography (EEG), can now be found in earables \cite{roddiger2022sensing, azemi2023biosignal}, headphones \cite{getversus, neurable}, or fabric headbands \cite{muse, brainbit2025}, while indirect measures can be derived from other physiological data, such as heart rate variability (HRV), skin temperature, or breathing rate, as collected through wristbands or smart rings. The new form factors allow for long-term usage in everyday settings and hereby open the door to an increasing number of use cases. While their ability to accurately assess users' cognitive states and performance still lags \textit{far} behind medical-grade laboratory devices, they are establishing their presence on the market, claiming to estimate users' stress levels (e.g., apple watch \cite{applewatch}), concentration (e.g., Versus \cite{getversus}), focus (e.g., Neurosity \cite{neurosity} or Neurable \cite{neurable}), or our mental readiness for the day ahead (e.g., Oura \cite{oura}), or support users' overall cognitive well-being (e.g., meditation exercises using Muse \cite{muse}). These devices make us wonder: will we soon be able to track our inner states and mind just as easily as we count our steps? And what will be the consequences of it?

In contrast to physical activity tracking, tracking cognitive data is inherently more complicated. While a step can be counted, attention is intangible, difficult to measure in demand, and highly subjective and context-dependent. For example, we could ask how much attention the current task actually needs. Is high focus always the goal? And do all individuals have the same baseline? How do we account for neurodivergent individuals, their motivation for cognitive tracking, and the quantification of data into metrics?

Overall, CPI research is still in its infancy; however, CPI and consumer neurotechnologies are receiving increased attention in HCI and are likely to have a far-reaching impact not only on our research but also potentially on our lives (cf. \cite{schneegass2024broadening}). While it sounds intriguing to measure students' attention in class or to optimize one's stress level to avoid burnout, we must ensure that such technology is designed responsibly to prevent harm.

We already started this discussion during our \textbf{past SIG at CHI 2022, and our workshops at MobileHCI 2023 \cite{mHCIworkshop} and CHI 2024 \cite{wilson2024chi}} with the HCI community. However, these prior events primarily focused on building a community around the topic, discussing challenges with current CPI research and the devices used, and outlining a research agenda for the following years. 

Thus, we plan to use this year's workshop to go into depth, as we see the pressing need to extend our focus to incorporate the technology development of the past three years -- specifically, the effects of generative artificial intelligence (genAI). GenAI is advancing rapidly, offering new opportunities for CPI. Beyond its ability to improve data analysis and cognitive state classification, it could support users in making sense of complex cognitive data, provide personalized feedback, and even act as a personal cognitive assistant or companion -- the latter appears to be a trend we have increasingly seen researched in 2025 (cf. \cite{prolongedusage,persistentassistant, freire, xi2025liviaemotionawarearcompanion, gao2025onlinematellmbasedmultiagentcompanion, yang2025chatwisestrategyguidedchatbotenhancing}).

Despite its potential, we see several risks associated with AI use in CPI systems, which we want to incorporate into our discussions at CHI 2026, including bias in interpretation, loss of user agency, and the potential for manipulative or exploitative use.

\subsection{Open Questions \& Research Gap}
We summarize several gaps and challenges that are critical and relevant to the HCI community. Our workshop focuses on the following core questions:

\paragraph{What should we sense?} -- Which signal streams are the most meaningful for CPI and how can they be measured (e.g., through direct or indirect [behavioral, physiological, or contextual] data)? For aspects of cognition that are interesting for HCI but difficult to assess, such as mind-wandering, how could they be approximated or supported by other data streams? 

\paragraph{How should we present the data?} -- Prior research on personal informatics has shown that data needs to be conveyed in an understandable manner and create a meaningful tracking experience~\cite{epstein2015lived, niess2018supporting}. However, the inherent complexity of cognitive data leads to added challenges, such as the risk of oversimplifying complex processes, such as workload or presenting data out of context. Additionally, research on the design of meaningful cognitive metrics is sparse, yet urgently necessary. For example, how should a complex process, such as workload, be presented when there is no baseline, and it depends on the task whether a high or low level is desired \cite{wilson2018mental, midha2022ethical}? Or the other way around - how do we present data from one underlying data stream, such as HR and HRV, that can give an indication to multiple cognitive facets \cite{nicolini2024heart, kim2018stress}, including executive functions, memory, and stress? And what do we have to consider when designing for neurodivergent populations (cf. \cite{Mitrevska2025Neurofeedback})?

\paragraph{How do we keep the user in the loop?} -- A frequently reported problem with cognitive tracking devices is a discrepancy between objective data and subjective experience \cite{dritsa2025data}. This stems not only from an inferior sensing or classification accuracy, but also from skewed self-perceptions that need to be mitigated through guided reflection of CPI data. It remains to be discussed how this can be achieved, what opportunities for feedback, data annotation, correction, and deletion these devices should offer.

\paragraph{How should AI be integrated and moderated?} 
One of the strengths of large language models (LLMs) lies in handling big data and making sense out of it by discovering unique patterns. This is a largely beneficial integration for handling personal psycho-physiological data, supporting a higher goal of less generalization in the interpretation and more tailoring to the unique individual \cite{cameron_llms, dongre2024integratingphysiologicaldatalarge}. On a physiological level, data such as burned calories, HRV, steps, etc., can be generalizable between individuals that belong to the same age group, weight, and physical readiness \cite{Lischke2017,hrv}. However, the complexity of interpretation rises when cognitive data is in question. There is a higher variability between individuals, since cognitive states are inferred from behavior, mood, environment, and task type \cite{Brysbaert2024, Boogert2018-mv}. For instance, two individuals can have the same reaction time but different levels of mental workload. The multidimensionality of the data presents a challenge in interpreting it with classical statistical models or rule-based systems. Here, LLMs can help merge multiple data streams - (psycho)physiological measures, contextual factors, and user baselines - and synthesize them into coherent, human-readable insights. However, the implementation leaves many open questions: How can LLMs be put into practice? How to shape the feedback, user control, and transparency?

\paragraph{How can we mitigate ethical challenges?} -- 
The topic of CPI is closely connected to neuroethics, which is slowly becoming more discussed in the HCI community \cite{martinez2022understanding, berger2023ai, shein2022}, neurosurveillance \cite{muhl2023neurosurveillance}, neurorights \cite{ligthart2023minding}, and with many implications around data privacy, inclusion, and risks of technology misuse.  
It yet has to be defined, even by policies such as the GDPR, what ``mental'' data is, how much it tells us about our actual health, how the data can be used, and what needs to be limited. Furthermore, many open questions remain, such as in which contexts can what data be responsibly collected? How can we foster inclusive design by respectfully designing for inter-personal cognitive variance and neurodiversity \cite{burtscher2024neurodivergence}? And what new implications might emerge from using AI for data analysis - for example, will we be able to read people's minds \cite{rainey2024rights, rainey2020brain}?


\subsection{Why Now: The Right Time for a New CPI Workshop}
We see three current movements that led us to consider CHI 2026 as a perfect venue for our next instance of this workshop:
\begin{itemize}
    \item The \textbf{increasing number of consumer wearables} on the market with a form factor and usability to allow for everyday cognitive tracking even for non-experts.
    \item The \textbf{integration of (gen) AI} for data processing and interpretation, but also as a more complex intervention, such as through the concept of cognitive companions.
    \item The increasing availability of neurotechnologies on the consumer market has led to \textbf{societal concerns} and calls for more research and \textbf{stricter regulation} of the technology~\cite{ibc2020neuroethics}.
\end{itemize}

\section{Organizers}
Our organizing team brings together expertise from various HCI domains and beyond as follows (in alphabetical order):


\textbf{Francesco Chiossi} is a Senior researcher in Human–Computer Interaction at LMU Munich, with a background in cognitive science and a PhD in computer science. His work bridges multimodal computing, adaptive Mixed Reality interfaces, and AI-driven personalization, with a focus on human-centered design and trustworthy user experiences. More recently, his research has examined how large language models and AI-mediated interactions influence user agency, privacy, and interaction design in mobile and ubiquitous contexts. \textit{Francesco will lead from the physiological computing, adaptive system, and AI-interaction perspective.}

\textbf{Anna L. Cox} is a professor of HCI at the UCL Interaction Centre, at University College London. Anna's research focuses on the use of digital technology to support productivity, work-life balance, and workplace wellbeing. \textit{Anna will lead from the perspective of the future of work.}

\textbf{Dimitra Dritsa} is a Postdoctoral researcher at Delft University of Technology. Her work focuses on the intersection of physiological sensing, data sensemaking, and information visualisation. She brings experience in designing algorithms and tools that enhance the sensemaking process of physiological data, such as electrodermal activity. Currently, she explores topics ranging from how information from wearables should be best presented to understanding the effects of differences between objective and subjective data on user experience. \textit{Dimitra will lead from the perspective of physiological computing and data sensemaking.}

\textbf{Teodora Mitrevska} is a PhD candidate at LMU Munich. She holds a M.Sc. degree in Human-Computer Interaction from LMU Munich. Her work focuses on the usage of electrophysiological signals as a form of implicit feedback in adaptive systems. She brings expertise in classifying human perception of visual stimuli and user-intent modeling for human-in-the-loop systems (HITL). She also takes an interest in consumer devices for cognitive tracking and their integration in everyday life.\textit{Teodora will lead from the perspective of quantification of electrophysiological data and AI for sense-making.}

\textbf{Stephen Rainey} is Assistant Professor in Philosophy and Technology at Delft University of Technology, specialising in neuroethics and neurophilosophy as part of a wider philosophy and technology agenda. He is currently exploring intersections between neurotechnologies and Artificial Intelligence, especially Large Language Models and the prospect of mind-reading technology. He applies research findings in policy advice, working with committees of the European Commission and the WHO. \textit{Stephen will lead this workshop from the perspective of neuroethics and neurorights.}

\textbf{Christina Schneegass} (main contact) is an assistant professor for Cognition \& Design at Delft University of Technology, who focuses in her research on the design of interfaces and interaction techniques with personal cognitive data from an end-user perspective, as well as using sensing data to augment users’ cognitive processes. \textit{Christina will lead from the perspective of user-centred design for cognition-aware systems and cognitive augmentation.}

\textbf{Max L. Wilson} is an associate professor at the University of Nottingham, focused on evaluating the mental workload involved in completing work tasks and created by differences in user interfaces, using qualitative investigations and quantitative studies using fNIRS. Max has also worked on brain-controlled movies that have toured around the world using consumer brain devices. Max is also a member of the IEEE Brain NeuroEthics Committee. \textit{Max will lead from the perspective of reflecting on CPI data.}

\section{Format \& Activities}
Through hosting our prior SIG and workshops, we managed to create a small community around researchers and practitioners interested in the topic of CPI. We already have an existing Slack\footnote{\href{https://join.slack.com/t/cog-pers-informatics/shared_invite/zt-14gt4zhte-Zq0U6ueTdS0X5qeWPDnhWQ}{CPI Community Slack Server}} channel with 90 registered members and a medium blog\footnote{\href{https://medium.com/@cogpi}{Medium Blog Cognitive Personal Informatics}} we use to disseminate any related content, such as new publications, opinion articles, and calls for special issues or workshops. We plan to use these channels to advertise our workshop. Additionally, we hope to further extend our community by advertising the workshop through our institutional channels, at upcoming HCI conferences (e.g., IUI26, MUM25), and at other relevant events (e.g., Dagstuhl seminar\footnote{\href{https://www.dagstuhl.de/en/seminars/seminar-calendar/seminar-details/25422}{Dagstuhl Seminar on Cognitive Sensing \& Interaction, October 2025}}). We will also recruit participants through CfP mailing lists and social media channels, ensuring a broad reach and facilitating inclusivity, beyond North American and European subcommunities.

\paragraph{Submissions \& Selection Process}
Instead of a classical submission of empirical or position papers, we will provide a template of guiding questions that participants should answer and submit. These questions will support prospective participants in reflecting on their work in relation to CPI, start thinking about our topic, and create a foundation for the discussion during the workshop. Questions include, among others:

\begin{itemize}
    \item What is your motivation to submit to this workshop - what draws you to the topic of CPI?
    \item What do you consider the three main challenges of CPI that HCI research needs to address?
    \item Where do you see risks of AI when used for CPI data analysis, interpretation, or mediating the interaction between data and user in any other way?
    \item What will CPI look like 10-20 years from now: Describe a positive scenario (I hope this will be possible) and a negative scenario (I fear this might happen). 
\end{itemize}

The submissions will be handed in via our website\footnote{\href{https://brain-data-uon.gitlab.io/events/chi26-workshop.html}{CHI 2026 CPI Workshop Website}} and reviewed by the organizers and, if needed, invited external experts. We will accept submissions based on their value to stimulate discussion, contribute to answering the open questions described above, or enrich the research on CPI in general.  

Based on our experience with prior workshops, we expect around 30 submissions and an acceptance rate of above 90\%, only rejecting submissions that are clearly out of scope for our workshop.

\paragraph{Pre-Workshop Engagement}
During the workshop, we will collect notes, ideas, feedback, and any other comments via our Miro board\footnote{\href{https://miro.com/app/board/uXjVJEmOMoI=/?share_link_id=837062656149}{Workshop Miro Board}}. This Miro board will already be used to facilitate pre-workshop engagement of the participants with the workshop and topic, as well as with each other. We will have a dedicated introduction corner and will encourage people to share something about themselves and get to know the other attendees, thereby hoping to limit the time needed for introduction during the workshop. Furthermore, we will already present participants' submissions, anonymized, grouped by question, and clustered, for everyone to reflect on. We are aware that engagement prior to the workshop is often difficult to achieve, and thus, we limit it to the minimum necessary to allow us to jump right into the discussion during the workshop.

\paragraph{At the Conference}
We will conduct two sessions of 75-90 minutes, i.e., one afternoon of the conference program. 
\begin{itemize}
\item \textbf{Opening --} The organizers will open the workshop by briefly introducing themselves, presenting the topic, and outlining the agenda for the day.
    \item \textbf{Session 1: Group Discussion on Shared Interests and Mapping Challenges --} Since the participants already introduced themselves via the Miro board, we will do only a short introduction session at each discussion table so that group members can situate the knowledge and perspectives the other participants are offering. Afterward, we will jump directly into the first breakout session, in which small groups will work on discussing and mapping challenges for different CPI aspects. We will use participants' submissions as a basis to extend on.
    \item \textbf{Facilitated Coffee Break --} To encourage informal discussions during the coffee break, we will ask our participants to rank the challenges identified by every group in session 1 using stickers on the Miro board.
    \item \textbf{Session 2: Defining the Research Agenda and Road Map --} Building on the results of session 1 and the output of our previous workshop, this session will sketch a research agenda and road map for the most critical challenges. Every group will work on the challenges ranked highest on the Miro board, analyse them by timeline, impact, stakeholders, and other relevant factors, and define concrete next steps for research to address them.
    \item \textbf{Closing --} We will end our workshop by briefly summarizing the day and discussing our publication and post-workshop plans to facilitate future opportunities to work together and extend the community. The Miro board will stay available during and after the sessions to be used for our community for collaborative ideation and critical discussions.
\end{itemize}

Along with the guidelines of CHI 2026, we do not plan for hybrid attendance, but will allow authors who are unable to travel to record a video that will be played to the in-person audience, as well as full access to the Miro board.

\paragraph{After the Workshop \& Expected Outcomes}
As main outcomes, we expect to distill from our sessions a \textbf{set of urgent key challenges} related to our topics of interest within the CPI domain, as well as a \textbf{research agenda and roadmap} to address them. We hope to compile the challenges and road map into a research publication and synthesize meta-level findings from our workshop for our community in a Medium blog post. 
On the level of individual research projects addressing the challenges, we will discuss and facilitate opportunities for collaboration among our participants during the workshop. Other potential follow-up activities will also be discussed, such as, for example:
\begin{itemize}
    \item Future workshops at CHI or other HCI conferences
    \item A special issue or edited book on CPI
    \item A \href{https://www.dagstuhl.de/de/seminars/dagstuhl-seminars}{Dagstuhl seminar} on (cognitive) personal informatics
\end{itemize}

\subsection{Accessibility}
Our goal is to deliver a fully accessible and inclusive workshop experience. Thus, we encourage potential participants to ensure that their submissions follow SIGCHI's Accessibility guidelines to help organizers, external reviewers, and fellow future participants engage with their submission. Prior to the workshop, we will provide participants with an opportunity to anonymously indicate any accessibility needs or preferences, and we will make every effort to accommodate those requests.

\section{Call for Participation}
This workshop explores the future of cognitive personal informatics and spans one afternoon (2x90-minute sessions). Looking beyond classifying cognitive states, the goal of this workshop is to examine \textit{how we can design people's interaction with consumer neurotechnologies in ways that are meaningful, ethical, and empowering, supporting reflection, agency, and well-being in everyday life}. Especially in a world where wearable technology is beginning to estimate stress, and consumer neurotechnology is available at a low cost.

To participate in our workshop, we invite people to fill in a template with guiding questions on their research's relationship to CPI, the challenges they see for the future, as well as scenarios they envision and/or are interested in working on. Any person who wishes to attend the workshop needs to submit this form. 

Submissions will be reviewed for how well the researchers' interests align with the goals of the workshop and to what extent the answers to the guiding questions provoke discussion and contribute to a better understanding of CPI. We consider research interests on physiologically-driven interaction and cognitive state classification out of scope. Submissions should follow our template and be handed in via \href{https://brain-data-uon.gitlab.io/events/chi26-workshop.html}{[our website]}. The workshop will be in-person only.


\bibliographystyle{ACM-Reference-Format}
\bibliography{sample-base}

\end{document}